# A novel convergence enhancement method based on Online Dimension Reduction Optimization

Wenbo Cao,[1] Yilang Liu,[2] Xianglin Shan,[3] Chuanqiang Gao,[4] and Weiwei Zhang[5*]
*School of aeronautics, Northwestern Polytechnical University, Xi'an 710072, China.*

Corresponding Author: Weiwei Zhang, aeroelastic@nwpu.edu.cn.

**Abstract:** Iterative steady-state solvers are widely used in computational fluid dynamics. Unfortunately, it is difficult to obtain steady-state solution for unstable problem caused by physical instability and numerical instability. Optimization is a better choice for solving unstable problem because steady-state solution is always the extreme point of optimization regardless of whether the problem is unstable or ill-conditioned, but it is difficult to solve partial differential equations (PDEs) due to too many optimization variables. In this study, we propose an Online Dimension Reduction Optimization (ODRO) method to enhance the convergence of the traditional iterative method to obtain the steady-state solution of unstable problem. This method performs proper orthogonal decomposition (POD) on the snapshots collected from a few iteration steps, optimizes PDE residual in the POD subspace to get a solution with lower residual, and then continues to iterate with the optimized solution as the initial value, repeating the above three steps until the residual converges. Several typical cases show that the proposed method can efficiently calculate the steady-state solution of unstable problem with both the high efficiency and robustness of the iterative method and the good convergence of the optimization method. In addition, this method is easy to implement in almost any iterative solver with minimal code modification.

**Keywords:** Computational fluid dynamics; Steady-state; Optimization; Physical instability; Numerical instability.

# 1. Introduction

COMPUTATIONAL fluid dynamics (CFD) has developed rapidly in the past 40 years, and has become an important means of theoretical research of fluid mechanics and engineering aerodynamic analysis and design. Iterative steady-state solvers are the primary workhorses to obtain the steady-state solution in CFD application. However, they cannot converge towards the steady-state solution for unstable problem caused by physical instability and numerical instability. For the inherently unstable flow with global instability modes [1] (physical instability), the lack of convergence most critically affects linearized flow analyses. The unstable steady-state solution $Q$ that satisfies the PDE residual $R(Q) = 0$ to machine precision is a prerequisite for the reliable linearized flow analysis [2] such as adjoint [3], instability analyses [4], resolvent analyses [5], and so on. Substituting a steady-state solution with a semi-converged or a time-averaged flow solution usually results in significant uncertainty [6, 7] and severely undermines the reliability of the analysis tool [8]. Moreover, the unstable steady-state solution is also very important for flow control, which is usually adopted as a design target [9]. In addition to physical instability, CFD also faces a variety of numerical instability problems that may lead to non-convergence, which may be caused by a large CFL number, insufficient dissipation, poor grid quality, non-robust high-order numerical schemes and turbulence models [10, 11] and so on. Although such convergence difficulties are commonly encountered in practice, they are seldom discussed in the literature [12, 13]. The lack of convergence of iterative solver limits the practical application of CFD.

For the unstable problem, the Newton's method [14] is a classical method to calculate the unstable steady-state solution. It is known for its quadratic convergence near the steady-state solution. Although it has been used to obtain the steady-state solution of some problems [8, 15, 16], the method may has severe practical limitations due to the sensitivity to the initial value and the computational cost for large and strongly nonlinear systems [17, 18]. For challenging flow problem with many complicated bifurcations at high Reynolds number, the Newton's method must be coupled with a continuation method [19]. In addition, there are still many CFD solvers that do not integrate Newton's method, and this extension is difficult. The Selective Frequency Damping (SFD) [1, 19] is a more popular method for computing unstable steady-state solution, which tries to reach the steady-state solution of an unsteady system by damping unstable temporal frequencies. The SFD method is easy to implement and has been successfully applied to the confined separated flow [20], the flow around a sphere [21], the turbulent separated flow around an airfoil at stall [22]. However, the parameters of this method depend on a frequency estimation of the globally unstable modes, and inappropriate parameters may result in non-convergence or computationally infeasible in practice. Several different methods [17, 23, 24] have been developed for adaptively determining the parameters of the SFD method, but it is still difficult for this method to obtain unstable steady-state solution for the unstable system containing multiple unstable modes [23]. BoostConv [25] is another method with small computational cost. It utilizes Krylov subspaces spanning the residual history to modify the residual at the current iteration step so that the residual at the next step is effectively reduced. This method is successfully used to accelerate convergence and obtain the steady-state solution for laminar flow. It is also improved later and applied to turbulent flows [26].

Although they have been widely used to calculate the steady-state solution of inherently unstable flow, as mentioned above, these methods are not robust enough and are seldom applied to solve numerical instability problem. We consider a novel method to solve the unstable problem. Noted that if the PDE is solved by optimization method, the norm of the residual is the objective function in the optimization and the residual of the steady-state solution is zero, so the steady-state solution is always the extreme point in the optimization regardless of whether the problem is unstable or ill-conditioned. Therefore, the optimization method is easier to solve the unstable problem than the iterative method. The optimization can be easily used to calculate the fixed points of ordinary differential equations, but it is difficult to solve partial differential equation directly because discrete partial differential equation is extremely high-dimensional in engineering. In this study, we perform POD on snapshots and optimize PDE residual in the POD subspace. By alternately using the iterative method and the optimization method to solve the PDE, this method can obtain the steady-state solution of unsteady flow efficiently and robustly. The rest of the paper is organized as follows. The ODRO method is introduced in section 2. Several typical cases are used to verify the effectiveness and parameter sensitivity of the method in Section 3. Finally, conclusions are drawn in Section 4.

## 2. Numerical Method

**2.1 Governing equation**

In this study, we perform numerical simulations using an in-house hybrid unstructured CFD code which solves the Reynolds-averaged Navier–Stokes (RANS) equations using a second-order cell-centered finite volume approach. The integral form of the two-dimensional compressible RANS equations for a cell of volume $\Omega$ limited by a surface $S$ and with an outer normal vector $\boldsymbol{n}$ can be written as follows

$$\frac{\partial}{\partial t}\iiint_{\Omega} \boldsymbol{Q} d\Omega + \iint_{s} \boldsymbol{F}(\boldsymbol{Q}) \cdot \boldsymbol{n} dS = \iint_{s} \boldsymbol{G}(\boldsymbol{Q}) \cdot \boldsymbol{n} dS \qquad (1)$$

where $Q$ is the vector of conservation variables $Q = [\rho \quad \rho u \quad \rho v \quad \rho E]$. $\rho$ is the density. $u, v$ are respectively the $x$-wise and $y$-wise components of the velocity vector of the flow. $E$ denotes the specific total energy. $F$ is the inviscid flux and $G$ is viscous flux with respect to molecular viscosity $\mu$ and turbulent viscosity $\mu_t$, which are calculated by Sutherland's law and Spalart-Allmaras (SA) turbulence model [27], respectively.

## 2.2 Proper orthogonal decomposition

POD is a linear method for establishing an optimal basis, or modal decomposition, of an ensemble of continuous or discrete functions. Detailed derivations of the POD and its properties are available elsewhere [28, 29] and not repeated herein. In this method, the snapshots $X = [u_1, u_2, \cdots, u_N]$ calculated by a few iteration steps of CFD solver are used to obtain the POD modes $\Phi$ by singular value decomposition.

$$X' = X - \bar{X} \quad (2)$$

$$X' = \Phi \Sigma V^* \quad (3)$$

where $u$ is the flow variables, $\bar{X}$ is the mean of the snapshots $X$, $\Sigma = diag(\lambda_i)$ is a diagonal matrix with the singular values of $X'$, $\Phi$ is a matrix with the left singular vectors and also the POD modes, and $V$ is a matrix with the right singular vectors. In practice, fewer than $N$ modes are retained to simulate system behavior.

In the POD subspace, the flow variables $u$ will be expressed as

$$u = \Phi \xi + \bar{X} \quad (4)$$

where $\xi$ is a vector of mode coefficients or mode amplitudes.

## 2.3 Optimization for solving PDE

In the POD subspace, the root mean square residual of steady RANS equations can be expressed as

$$R_{total}(\xi) = \sqrt{\frac{1}{n} \sum_{m=1}^{n} \|R_m(\xi)\|_2} \quad (5)$$

where $R_m(\xi)$ is the residual of the $m$th grid, and $n$ is the total number of grid elements. The $R_{total}(\xi)$ can be obtained simply by calling the iterative solver. Then the optimization can be implemented simply by calling the open-source optimization library such as *scipy* [30] to find the target solution with $R_{total}(\xi)$ minimized in the POD subspace. Comparing with the iterative method, this method is more flexible and suitable for both explicit and implicit discrete schemes, and can overcome the numerical instability of iterative method [31, 32]. In practice, the mode coefficients corresponding to the last snapshot can be used as the initial value of the optimization parameters to accelerate the optimization process.

### 2.4 Online Dimension Reduction Optimization

Summarizing the parts introduced above, the proposed method consists of the following steps:

(1) Perform a few iteration steps of CFD solver with the initial value $u_0$, and save the snapshots of the flow variables during the iterative stage.

(2) Operate POD analysis on snapshots to calculate the POD modes $\Phi$.

(3) Minimize the residual $R_{total}(\xi)$ using optimization to obtain a lower-residual solution in the POD subspace, then update $u_0$.

(4) Repeat steps (1-3) until the residual converges.

The parameters in this method include the number of snapshots $N$, the snapshot storage interval $K$, and the number of POD modes $O$. The *Nelder-Mead* algorithm [33, 34] is recommended for optimization, which is one of the best known algorithms for multidimensional unconstrained optimization without derivatives and highly efficient for low-dimensional problem. In practice, the maximum times of calling CFD to evaluate the objective function $R_{total}(\xi)$ in each optimization process is limited to $N \times K/10$ to ensure that the computational cost of optimization only accounts for about 1/10 of the total computational cost.

## 3. Results

In this section, several typical cases are used to verify the effectiveness and the parameter sensitivity of ODRO method. While simulating the following cases, implicit symmetric Gauss-Seidel scheme [35] is taken as the time-marching method, and Roe scheme [36] and second-order central scheme are used to calculate the inviscid fluxes and viscous fluxes, respectively.

### 3.1 The laminar flow past a cylinder

The first test case is the two-dimensional laminar flow past a cylinder at different Re, which is a classic example to obtain the unstable steady-state solution. When the Re is greater than the critical Reynolds number $Re_c \approx 47$, the flow is unstable and von Karman vortex streets are observable. Therefore, the purpose of the ODRO method is to suppress these oscillations and drive the solution towards its steady-state solution.

As shown in Fig. 1, the computational domain is composed of 67613 elements. Fig .2 shows the streamwise velocity and vorticity contour of the steady-state solution obtained by ODRO at $Re=100$. This flow field is identical to the one presented by Barkley [37]. A stability analysis, using the Arnoldi method, is performed with this steady-state as the "base flow". The two dominant (conjugate) eigenvalues are $\lambda_{1,2}=1.137749e^{\pm 0.758122i}$, which correspond to the values presented by Barkley [37]. The size of the separation zone shown in Fig.2a is also consistent with the results in [19] and reattachment is near 6.65m. Fig.3 shows the contour of steady-state solutions at $Re=200,300,500,700$. It shows that the flow becomes more and more unstable and the separation vortex tends to become larger with the increase of Reynolds number. To the best of author' knowledge, this is the first attempt of obtaining the steady-state solution with a maximum Reynolds number $Re=700$. Fig. 4a shows the convergence histories of the mean density residuals against iteration numbers, and it shows that the convergence gets slower as Re increases. Nevertheless, the residuals eventually decrease by more than 9 orders of magnitude at all test Reynolds numbers. The Fig. 4b is the local magnification of the convergence residual at $Re=700$, and it shows that the residual drops steeply every 400 steps and always rises during the interval, which verifies that the traditional iterative method always naturally deviates from the unstable steady-state solution as mentioned above and the optimization is always effective so that a solution with a smaller residual can be obtained. Finally, the ODRO method still maintains a global drop in residual.

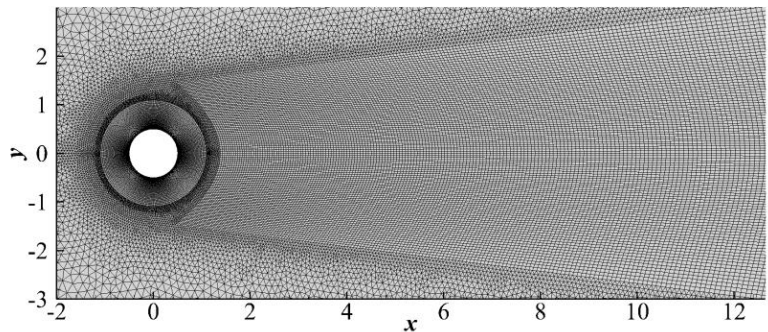

**Fig.1 The mesh of the flow past a cylinder**

**(a)**                                        **(b)**

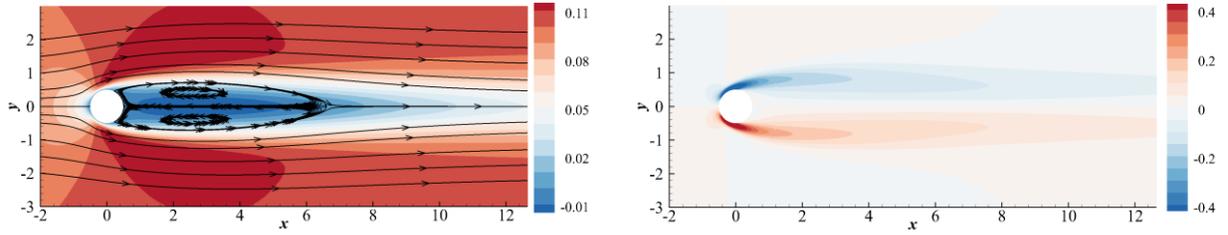

**Fig. 2 The unstable steady-state solution of the flow past a cylinder at Re=100. (a) The streamwise flow velocity contour. (b) The Vorticity contour.**

(a) (b)

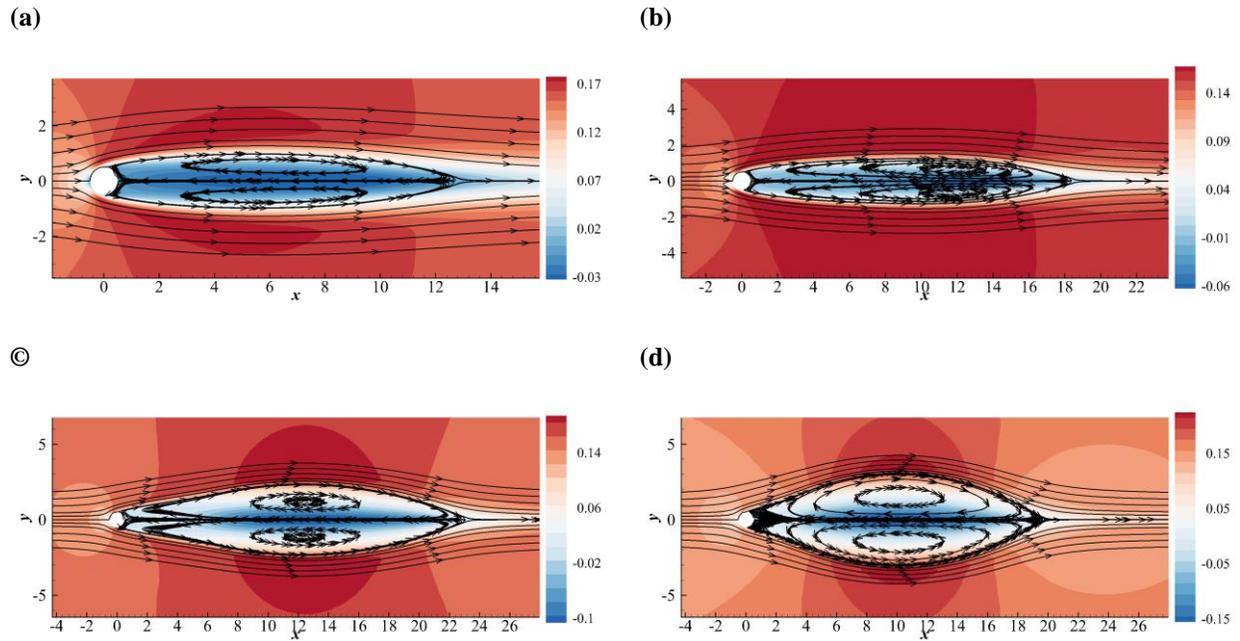

(c) (d)

**Fig. 3 The streamwise flow velocity contour of the flow past a cylinder at (a) Re=200, (b) Re=300, (c) Re=500, (d)Re=700.**

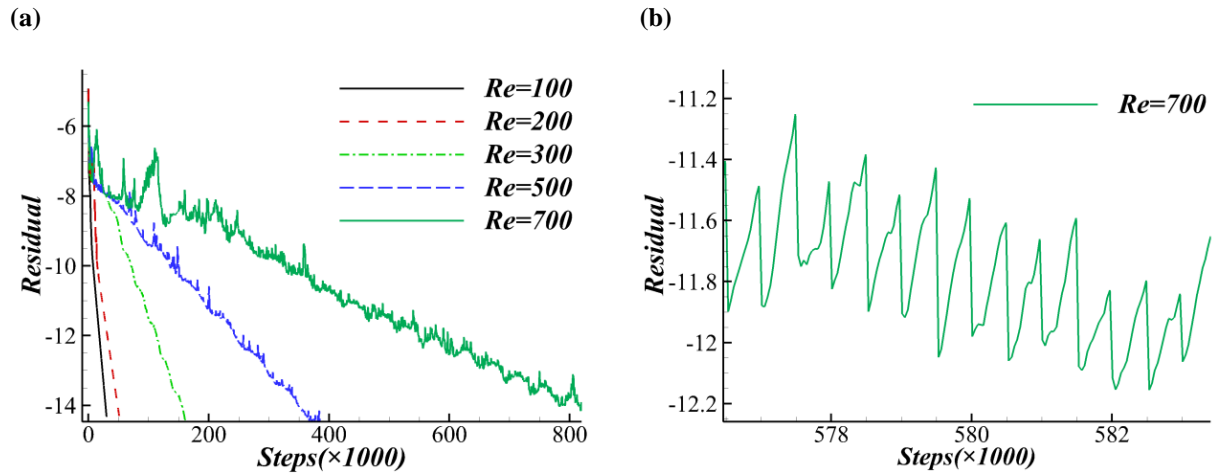

**Fig. 4 The convergence histories against iteration numbers. (a) Different Re. (b) A local magnification at** $Re = 700$.

In this case, we also tested the effect of the ODRO method for different parameters at $Re = 700$, which is the most unstable and stringent case of all the cases we have tested. As shown in the Fig. 5, the ODRO method has consistent effect in a wide range of parameter spaces as long as the call interval $N \times K$ of dimension reduction optimization is not too large. In particular, it shows that even if only 3 snapshots and 2 POD modes are considered, the unstable steady-state solution can be obtained by this method, which greatly reduces the memory requirements of this method. In all the following cases, the parameters in the ODRO method are set as $N = 5, K = 80, O = 5$. The runtime for the flow simulation at $Re = 100$ with $N = 5, K = 80, O = 5$ is recorded. The total runtime is 1373s, and the runtime for dimension reduction optimization is 136s. This indicates that optimization has little additional computational cost.

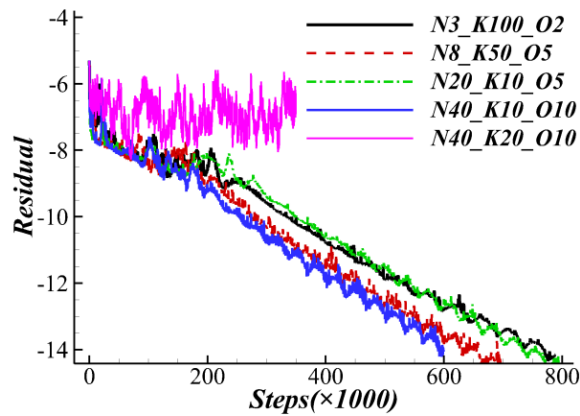

**Fig. 5 Convergence histories of different parameters.**

## 3.2 Transonic buffet flow

The second test case is the transonic buffet flow around NACA0012 airfoil. The proposed method is applied to the problem of transonic buffet flow on a two-dimensional airfoil as a validation study. The motivation for the choice of this problem comes from the increasing interest in the shock wave/boundary layer interaction in transonic flow, the control of which is also a challenging task in aerospace science and engineering. Transonic buffet is a phenomenon of aerodynamic instability at a certain combination of Mach number and mean angle of attack [38, 39]. The strong buffet unsteadiness, characterized by periodic low-frequency and large-amplitude shock oscillations, results in lift and drag fluctuation which leads to structural fatigue of the aircraft or launch vehicle [40]. For the stability analysis and flow control of transonic buffet, it is necessary to obtain its unstable steady solution as "base flow".

A Flow control method has been applied to obtain the unstable steady-state solution of NACA0012 airfoil at $Re = 3e6, Ma = 0.7, \alpha = 5.5°$ [9]. The purpose of the ODRO method is to reproduce this result. The computational domain consists of 18520 elements, and the turbulence model applied in the computations is the SA model, which is consistent with [9]. The Fig. 6a shows the residual histories comparison between the initial iterative method and the ODRO method. It shows that the ODRO method can successfully obtain the steady-state solution of the flow. Fig. 6b shows the consistent pressure coefficient distribution obtained by the flow control method and the ODRO method. Fig. 7a and Fig. 7b are the steady-state solution obtained by the flow control method and the ODRO method, respectively, and they are consistent.

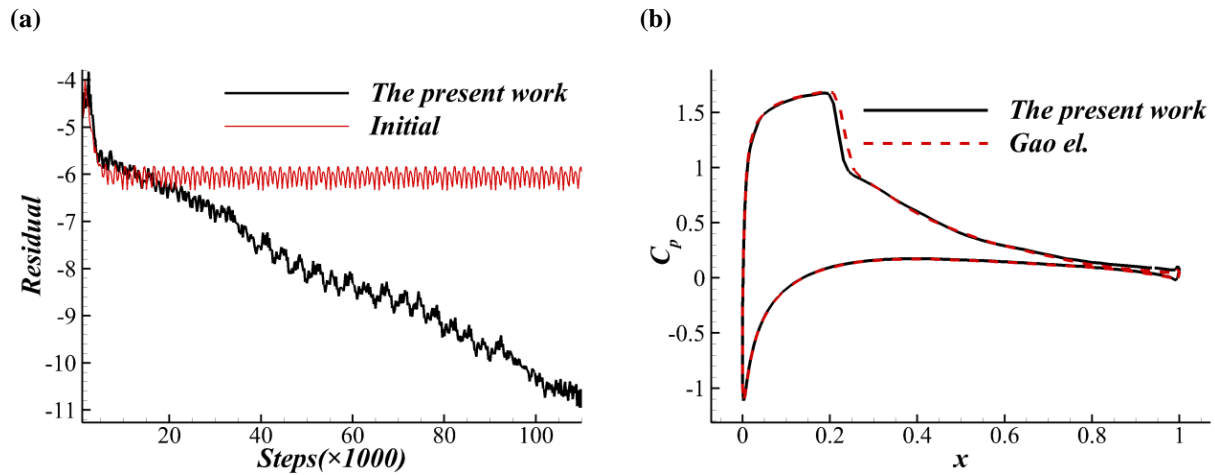

**Fig. 6 (a). Convergence histories of the initial method and the ODRO method. (b). Pressure coefficient distributions obtained from the flow control [9] and the ODRO method, respectively**

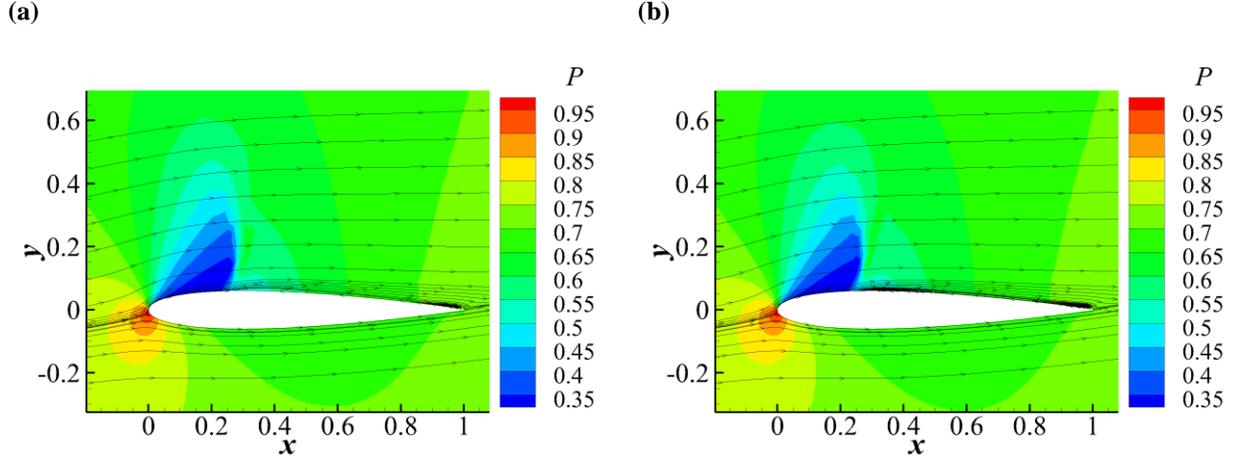

**Fig. 7 Pressure contours and streamline of the unstable steady-state solution obtained by (a) flow control method and (b) ODRO method.**

### 3.3 Turbulent separated flows around an airfoil at stall

The third test case is the turbulent stalling flow around a helicopter rotor blade airfoil OA209 at different angles of attack. Airfoil stall is a common complex in many military or civil aeronautical applications that severely affects the design of airplane wings, helicopter blades and engine turbine blades. It occurs at sufficiently high Reynolds number and high angle of attack. The laminar or turbulent boundary layer then separates from the airfoil, leading to a massive flow separation and an abrupt drop of lift that may cause a decrease of the aircraft's performance or even an uncontrolled fall. Understanding stall is an on-going research topic, and today its simulation and exploration still primarily relies on the classical RANS approach at high Reynolds number. However, it is often difficult to converge when the angle of attack approaches or exceeds the critical stall angle of attack. The lack of robustness of RANS approach is detrimental when analyzing the deficiency of turbulence models or predicting the onset of low-frequency fluctuations in turbulent flows with global stability analysis [22].

Newton's method has been applied to calculate the steady-state solution of OA209 airfoils at $Re = 1.8e6, Ma = 0.16$, and a continuation method is used to obtain the steady-state solution at high angle of attack because the Newton's method requires a good initial approximation [15]. An optional continuation method consists in incrementing the angle of attack $\alpha_1 = \alpha_0 + \Delta\alpha$ and computing the solution $Q(\alpha_1)$ with the Newton's method initialized by $Q(\alpha_0)$. Once it is determined, the solution $Q(\alpha_2)$ can be obtained from $Q(\alpha_1)$, and so on. This approach is feasible but inefficient. In contrast, the ODRO method can easily obtain the steady-state solution at high angle of attack with the free stream as the initial value.

The computational domain adopts CH-grid technology, which is composed of 124584 elements. The turbulence model applied in the computations is the SA turbulence model, which is consistent with [15]. The steady-state solution of the maximum angle of attack $\alpha_{max} = 22°$ is calculated in [15]. In this paper, we calculate the steady-state solution at $\alpha = 18°, 22°, 30°, 60°$. Fig. 8 shows the residual convergence histories of different angles of attack, which indicates that the ODRO method can efficiently obtain the steady-state solution with the free stream as the initial value in a large range of angles of attack. Fig. 9 is the contour of streamwise flow velocity at $\alpha = 22°$, which is identical to the one presented by Busquet [15].

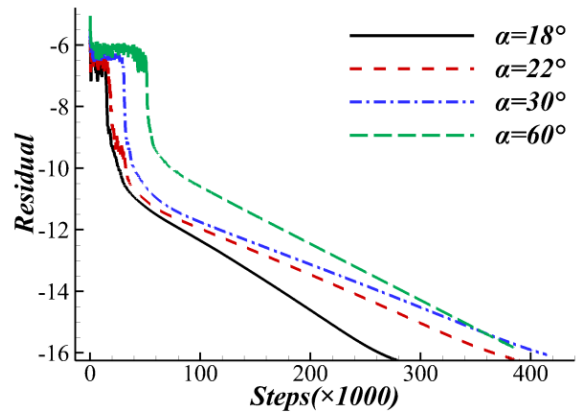

**Fig. 8 Convergence histories against different angles of attack.**

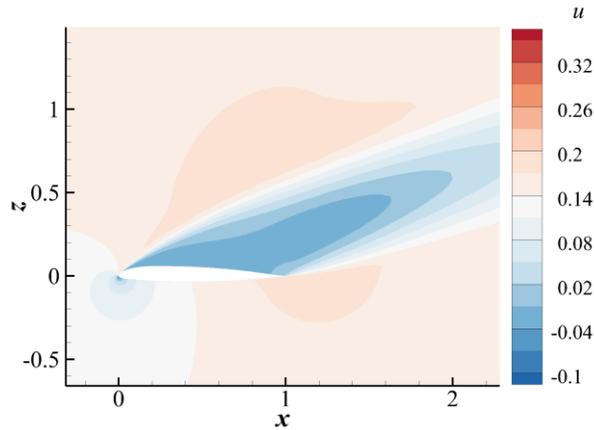

**Fig. 9 The streamwise flow velocity contour at $\alpha = 22°$**

### 3.4 Numerical instability caused by large CFL

As we all know, Courant-Friedrichs-Lewy (CFL) number significantly affects the convergence speed and numerical stability when the time-dependent method is used to calculate the steady solution. Even for implicit time discretization, CFL number may not be infinite due to many assumptions introduced, and the maximum number of CFL allowed for different flow types is often different. Therefore, the choice of CFL number still remains an open problem, which has not yet been solved [41]. As a result, in order to ensure numerical stability, the conservative CFL number is usually chosen, which leads to a slow convergence speed. In this subsection, Online Dimension Reduction Optimization will be used to overcome the numerical instability caused by a large CFL number, thus allowing significantly larger CFL number in the time-dependent approach to accelerate convergence.

The test case is the turbulent flow of S809 airfoil at $Re=2e6, Ma=0.15, \alpha=14.2°$. Fig. 10(a) and Fig. 10(b) respectively show the convergence histories of initial iteration method and ODRO method for different CFL numbers. Fig. 10(a) shows that, the convergence speed increases significantly with the increase of CFL number in a certain range, but the initial iteration method cannot converge when the CFL number reaches 20. While ODRO method can successfully converge at $CFL=20$ and $CFL=50$, and the convergence speed is significantly faster than the initial method at $CFL=5$. For larger CFL numbers ($CFL=200, 500$), the numerical scheme is very unstable so that CFD iteration diverges quickly even with the steady solution as the initial value, but the ODRO can still obtain the steady solution albeit with a slow convergence speed, which show its strong ability to overcome numerical instability. Fig. 11 shows the consistent pressure coefficient distribution and skin friction coefficient distribution obtained by the initial iterative method at $CFL=5$ and the ODRO method at $CFL=500$. It should be pointed out that the divergence is too fast due to numerical instability in CFD iteration stage at $CFL=500$, so the call interval $N \times K$ of ODRO is adjusted to 200 to ensure convergence.

(a)                                   (b)

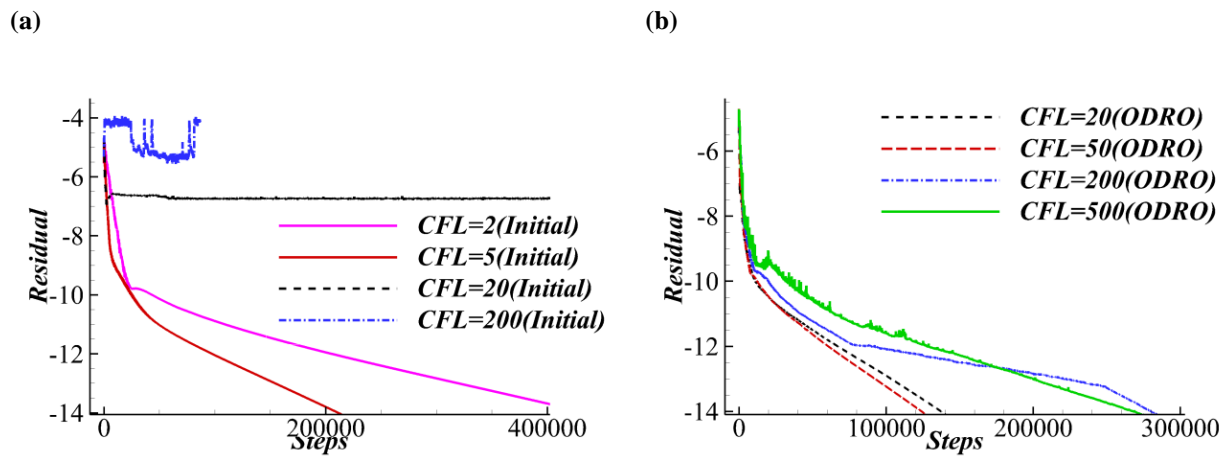

**Fig. 10 Convergence histories of different CFL numbers. (a) Initial iterative method. (b) ODRO method.**

(a)                                   (b)

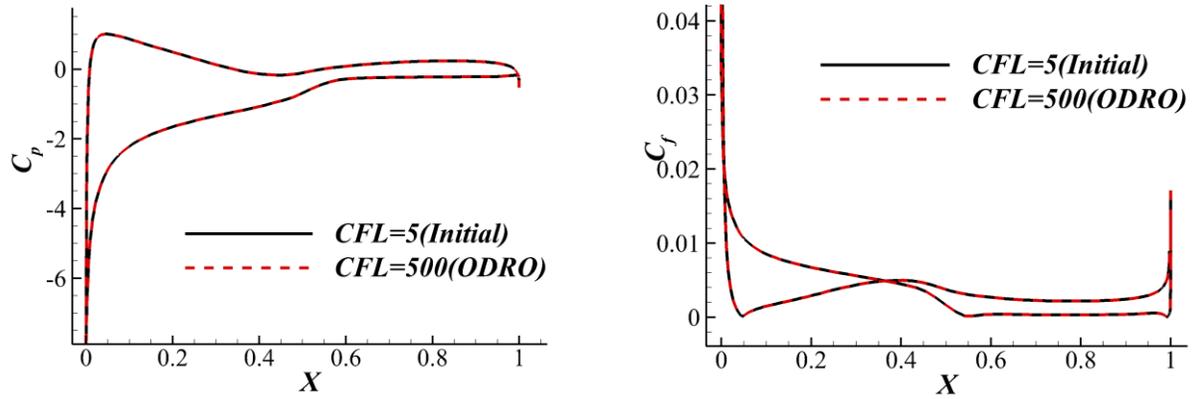

**Fig. 11 Comparison of surface (a) pressure coefficient and (b) skin friction coefficient.**

### 3.5 Numerical instability of non-robust turbulence model

RANS simulations are still the main method to study complex flows in engineering. However, traditional turbulence models cannot accurately predict flow fields with separations. Recently, machine learning methods provide an effective way to build new data-driven turbulence closure models trained on high-fidelity simulation databases [42-45]. Nevertheless, the traditional Reynolds-stress models and the recent machine learning turbulence models are faced with the serious numerical instability leading to non-convergence [10, 11], which extremely limits the development and application of these turbulence models. In this subsection, Online Dimension Reduction Optimization will be used to overcome the numerical instability caused by non-robust machine learning turbulence models.

Since the emphasis is how to overcome the numerical instability of machine learning turbulence models, concisely and generally, we use the data of the SA turbulence model to construct the mapping between mean flow features and the turbulent eddy viscosity with deep neural network (DNN), and the input features and architecture of DNN turbulence model of our previous work [11, 46] are adopted. Then numerical simulation is performed by substituting the SA turbulence model with DNN turbulence model. The test case is the turbulent flow of NACA0012 airfoil with 76356 grids at $Re=6e6, Ma=0.15$. Fig. 12 shows the convergence histories of the initial iterative method with DNN turbulence model at different angles of attack. It shows that when the angle of attack is greater than 10°, RANS equations cannot converge. While the convergence histories of the ODRO method is shown in Fig 13a, which indicates that RANS equations with DNN turbulence model converge to near machine zero at all angles of attack. The Fig. 13b is the local magnification of the convergence history at $\alpha=15°$, and it shows that the residual always rises in the iterative process due to the numerical instability of DNN turbulence model and drops steeply every 400 steps due to the good convergence of optimization in the POD subspace. Finally, the ODRO method maintains a global drop in residual. Fig. 14 shows the consistent pressure coefficient distribution and skin friction distribution obtained by SA turbulence model and DNN turbulence model.

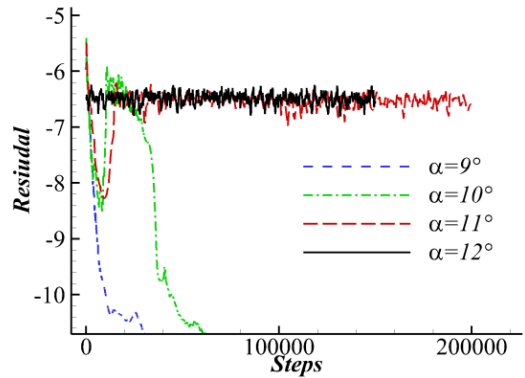

Fig. 12 Convergence histories of the initial method at different angles of attack.

(a) (b)

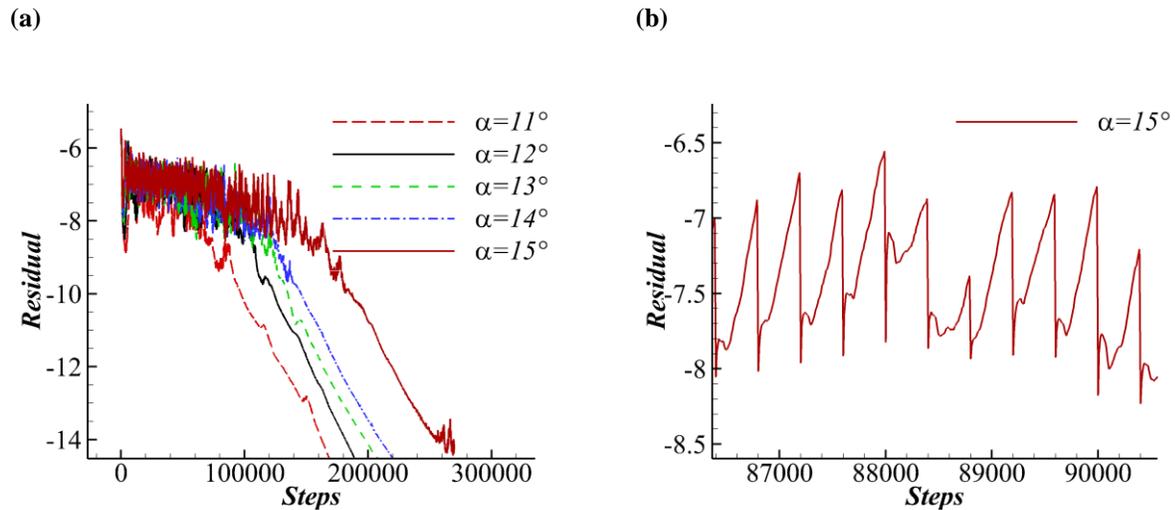

Fig. 13 Convergence histories of the ODRO method. (a) different angles of attack. (b) A local magnification at $\alpha = 15°$.

(a) (b)

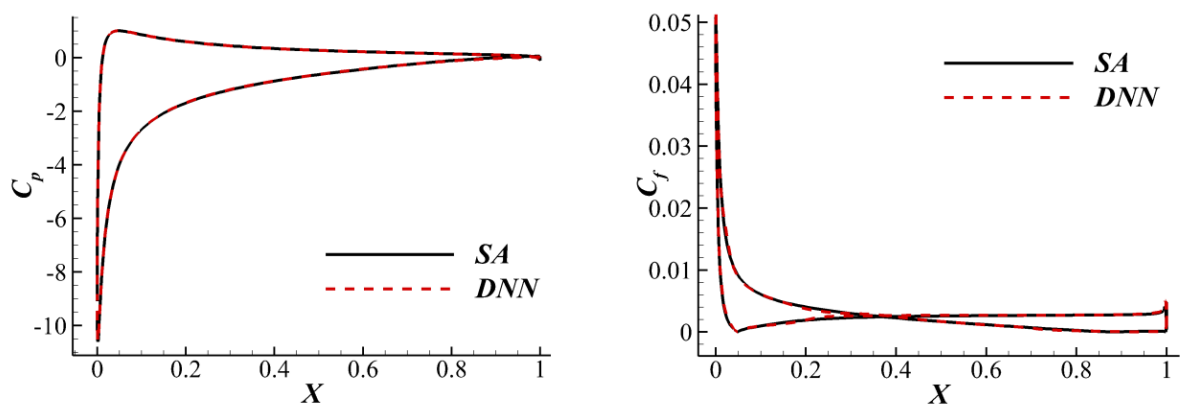

**Fig. 14 Comparison of surface (a) pressure coefficient and (b) skin friction coefficient between SA turbulence model and DNN turbulence model.**

## 4.　Conclusions

In this paper, Online Dimension Reduction Optimization method is proposed to enhance the convergence of the traditional iterative method to obtain the steady-state solution of unstable problem. This method uses iterative method and optimization method to solve the PDE alternately, which not only inherits the high efficiency and robustness of the iterative method, but also has the good convergence of the optimization method. The method is easy to implement in almost any iterative solver with minimal code modification. A simple demo is provided for readers' reference on *https://github.com/Cao-WenBo/Online-Dimension-Reduction-Optimization*. (The code will be uploaded after the paper is reviewed)

The method developed is applied to the steady-state solution computation for the laminar flow past a cylinder, the transonic buffet, the turbulent separated flow, and the numerical instability problems caused by large CFL number and non-robust turbulence model. These cases show that the ODRO method is robust and efficient, can always stably reaches machine-zero converged solutions, and is insensitive to its parameters and the initial flow fields. In particular, the method works even if only 3 snapshots and 2 POD modes are considered, which means that the method has lower memory requirements and additional computational cost for optimization. In addition, because the POD modes come from the snapshots of the iteration and are constantly updated, the numerical accuracy of the method still depends entirely on the iterative method.

We expect this method can efficiently and robustly obtain the unstable steady-state solutions with machine zero precision, thus allowing reliable linearized analysis such as adjoint, instability analyses, resolvent analyses, and so on. We also expect that this method can solve the numerical instability problems caused by a large CFL number, insufficient dissipation, poor grid quality, non-robust turbulence models, and so on, thus leading to a more robust CFD solver.